# Effect of Exchange-type Zero-bias Anomaly on Single Electron Tunnelling of Au Nanoparticles

Rui Xu[1], Yi Sun[1], Hui Yan[1], Ji-Yong Yang[1], Lin He[1,a,*], Jia-Cai Nie[1,b], and Yadong Li[2,c]

[1]Department of Physics, Beijing Normal University, 100875 Beijing, People's Republic of China

[2]Department of Chemistry, Tsinghua University, 100084 Beijing, People's Republic of China

Using cryogenic scanning tunnelling microscopy and scanning tunnelling spectroscopy we measured single electron tunnelling of isolated Au nanoparticles with 1.4 nm in radius. We observe that a gap $\Delta V \sim 2e/C$ ($C$ is the capacitance of the Au particle) around zero bias in the tunnelling conductance spectrum, followed by a series of discrete single electron tunnelling peaks with voltage widths of $E_C \sim e/C$ at both negative and positive bias. Experimental data are well explained by taking into account the effect of exchange interaction of electrons on the single electron tunnelling of Au nanoparticles. A tunnelling peak near zero-bias was suppressed by the exchange-type zero-bias anomaly, which results in the gap $\Delta V \sim 2E_C$.

Nanoparticles (NPs) are the subject of intense research at present, in part because they have opened up a new area of fundamental science and in part because of their long-term potential applications [1-3]. It is well-known that when put an individual nanoparticle (NP), isolated in an insulating barrier, between source and drain electrodes, then the NP acts as an island of electrons [4]. The tunnelling of electrons from the source through the particle to the drain can be inhibited at small bias voltages if the electrostatic energy $e^2/2C$ of a single excess electron on the particle is much larger than the thermal energy $k_BT$, where $C$ is the capacitance of the NP. Additionally, if the resistance $R_t$ of two tunnelling barriers between the NP and the two electrodes is much larger than quantum resistance $R_q = h/e^2$, which ensures that the wave function of an excess electron on the NP is localized there, the so-called Coulomb blockade (CB) could be observed. Since the pioneer experiment on CB of Fulton and Dolan [5], the single electron tunnelling behaviors of NP have attracted much attention [6-13]. Several groups explored unexpected phenomena of CB for ultrafine NP [14-16]. For example, it is found that the electronic wavefunctions of semiconductor quantum dots exhibit atomic-like symmetry and the semiconductor quantum dots can be treated as 'artificial atoms'. As a consequence, the separations of discrete single electron tunnelling for semiconductor quantum dots are determined by both the single electron charging energy and the discrete level spacings of the 'artificial atoms' [15]. These results indicate that the electronic properties of NP could influence itself the single electron tunnelling behavior.

In this Letter, we measured single electron tunnelling of isolated Au nanoparticles with 1.4 nm in radius by using cryogenic scanning tunnelling microscopy (STM) and scanning tunnelling spectroscopy (STS). A gap $\Delta V \sim 2e/C$ ($C$ is the capacitance of the Au particle) around zero bias, followed by a series of discrete single electron tunnelling peaks with voltage widths of $E_C \sim e/C$ at both negative and positive bias, is observed in the tunnelling conductance $dI/dV$ spectrum. The appearance of the gap $\Delta V \sim 2E_C$ and the magnitude of conductance peaks in the differential conductance curves are well explained by taking into account the effect of exchange interaction of electrons on the single electron tunnelling of Au NPs.

The colloidal gold solutions were prepared using oleic acid assisted solution method. The Au NPs were prepared with the previously reported procedure [17,18]. The obtained Au NPs passived by oil amine were redispersed in 35 ml n-hexane for further use. The size of the Au NPs is determined by transmission electron microscopy (TEM) which were carried out by a JEOL JEM-2100F microscope. The as-grown samples were dispersed to the copper grid which was coated with carbon membranes, then drying natural in air. Fig. 1(a) shows a typical TEM image of the Au NPs. The TEM image showed that the Au NPs have a distribution of sizes centered at radius of 1.42 ± 0.15 nm, as shown in Fig. 1(b). The two dimensional (2D) Au NPs array shows a local hexagonal-close-packed superlattice but lack long range order due to the size and shape distributions of the NPs.

In the STM investigations, the Au NPs were dispersed on Au (111) atomic flat substrate. The STM system was an ultrahigh vacuum four-probe scanning probe microscope from UNISOKU. All the STM and STS measurements were acquisition at liquid-nitrogen temperature in a constant-current scanning mode. The material of the STM tip is Pt (80%)/ Ir (20%) alloy. An asymmetric double-barrier tunnel junction (DBTJ) is formed when an STM tip is positioned above a ligand stabilized gold particle that was deposited on Au (111) substrate. A typical STM image of an isolated Au NP is shown in the inset of Fig. 2(a). There are several important paramters determined the electronic properties of the studied DBTJ: the capacitance (resistance) $C_S$ ($R_S$) and $C_D$ ($R_D$) of the particle-tip junction and substrate-particle junction respectively. These parameters are mainly determined by the thickness of the coated ligand ~ 2.5 nm, the distance between the STM tip and the particle ~ 1.0 nm, and the dielectric constant of the oil amine $\varepsilon_r$ ~ 2.67 [19]. The local electronic properties of a single Au NP were characterized by STS performed at 77 K by using standard lock-in techniques to obtain differential conductance ($dI/dV$-$V$). Fig. 2(a) and (b) show two typical STS curves obtained in the Au NPs. The main difference of the two curves is the bias dependent magnitude of the conductance peaks. Except this difference, the $dI/dV$-$V$ curves show a common feature that the voltage spacing between the first positive and negative tunnelling peaks $\Delta V$ is about twice of that of other neighboring peaks. The uniform periodic peaks at high voltage bias separated by $E_C \sim e/C$ should be attributed to CB behavior. By taking into account the capacitance of the Au NP (with assuming the radius of the Au NP is 1.42 nm), the spacing of the conductance peaks $E_C$ arising from the single electron tunnelling is estimated to be about 0.8 eV [20], which is consistent with our experimental result.

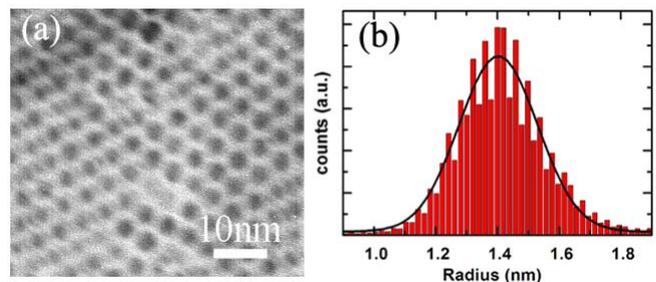

FIG. 1. (a) A typical TEM image of the Au NPs. (b) The size distribution of Au NPs shown in Fig. 1(a). The solid curve is the Gauss fit. It showed that the Au NPs have a distribution of sizes centered at radius of 1.42 ± 0.15 nm.

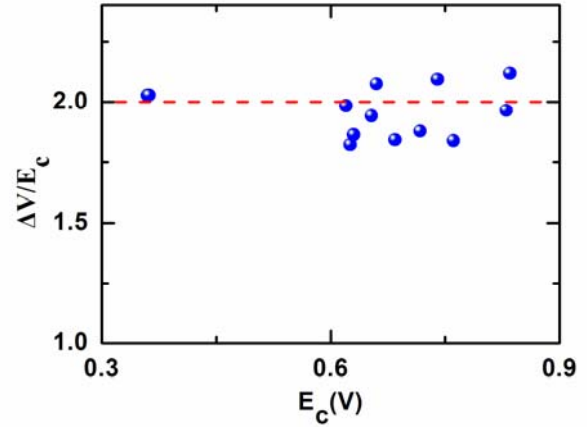

FIG. 3. The ratio of $\Delta V/E_C$ as a function of $E_C$ obtained in several different Au NPs.

The most striking and unexpected result of our experiment is the appearance of a gap $\Delta V$ around the zero bias in the STS curves of the Au NPs, which is unexpected within the orthodox CB theory [21,22]. According to the orthodox CB theory, the tunnelling peaks should show uniform voltage width of $E_C$. To eliminate the possible influence of the substrate and the STM tip, we carried out control experiments that we measured the differential conductance of the Au NPs on a highly ordered pyrolytic graphite (HOPG) substrate and by using W tips. Similar results are obtained. To explore the precise origin of the gap, we carried out STS measurements on more than ten different Au NPs on different substrates. Fig. 3 shows the ratio of $\Delta V/E_C$ as a function of $E_C$ obtained in our experiments. It is interesting to find that the gap $\Delta V$ is about twice of $E_C$ for different Au NPs. It indicates that we observe a superposition of the CB gap $E_C$ arising from the CB behavior of Au NPs and an additional gap, which also shows voltage width of $E_C \sim (\Delta V - E_C)$ around zero bias.

We now turn to understand the main experimental result, *i.e.*, the appearance of the gap around zero bias in the differential conductance curves. We should stress that the gap observed in our experiments is not the energy level spacing induced by quantum size effect. First, the energy level spacing arising from the quantum size effect can be simply estimated by $\Delta E = (1/3\pi^2 N)^{1/3} h^2/(4mL^2)$, where $L$ is the size of NP, $N$ is the number of electrons in NP, and $m$ is the effective mass of electrons [4]. For the Au NPs, the energy spacing is estimated as $\Delta E \sim 10$ meV, which is more than one order smaller than the experimental result ~ $eE_C$. Second, the the energy level spacing induced by quantum size effect should not only result in anomaly around zero bias and can not explain the result $\Delta V/E_C \sim 2$ observed in our experiments.

Chen, *et al.* have reported the electrochemical ensemble Coulomb staircase experiments on solutions of gold NPs of varied core sizes ranging from 0.55 to 0.95 nm in radius [14]. They observed that the ligand-protected Au NPs with radius larger than 0.67 nm show metallic behavior, whereas the Au NPs with radius smaller than 0.67 nm show molecular behavior with a substantial gap between the highest occupied and the lowest unoccupied orbitals (HOMO-LOMO) of 0.4 to 0.9 eV. The magnitude of the HOMO-LOMO gap is the same order as our experimental result. Theoretically, Walter, *et al.* have shown that ligand-protected ultrafine Au NPs behave as superatom complexes [23], similar as the semiconductor quantum dots. However, the size of the Au NPs in our experiment is much larger than the critical size ~ 0.67 nm in radius. Additionally, the gap between the HOMO-LOMO only depends on the size of Au NPs and has no obvious relation with the Coulomb blockade gap $E_C \sim e/C$ [14]. For example, Chen, *et al.* observed a HOMO-LOMO gap of 0.7 eV along with $E_C \sim 0.27$ eV for Au particles with 0.67 nm in radius and a HOMO-LOMO gap of 0.9 eV along with $E_C \sim 0.3$ eV for Au particles with 0.55 nm in radius. In our experiment, we observe $\Delta V/E_C \sim 2$ for all the Au NPs. Therefore, we can eliminate the HOMO-LOMO gap as the origin of the gap observed in our experiment.

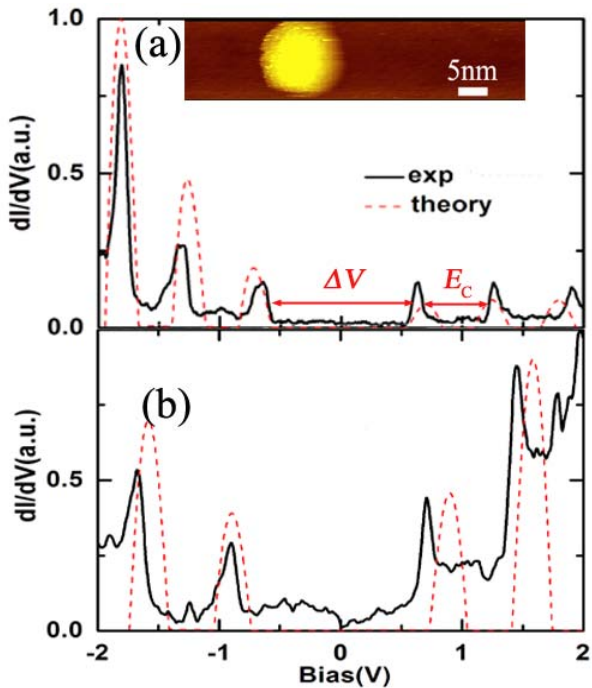

FIG. 2. (a) A typical $dI/dV$-$V$ curve (black solid curve) of Au NP measured at 77 K. The magnitude of the peaks decreases quickly with decreasing the bias from -2 V to 0 V and then keeps a constant with increasing the bias from 0 V to 2 V. The voltage spacing between the first positive and negative tunnelling peaks $\Delta V$ is about twice of that of other neighboring peaks $E_C \sim 0.65$ eV. The red dashed curve is the calculated result of Eq. (2) with parameters: $R_D$ = 7500 Ω, $R_S$ = 1.3×10$^6$ Ω, $C_S$ = 2.9×10$^{-19}$ F, and $C_D$ = 5.8×10$^{-19}$ F. The inset shows a typical STM image of an isolated Au NP. It was taken with a tunnelling current of 8.32 pA and a bias voltage of -1.18 V. (b) A typical $dI/dV$-$V$ curve of Au NPs measured at 77 K with the magnitude of the peaks almost symmetrical about zero bias (black solid curve). The red dashed curve is the calculated result of Eq. (2) with parameters: $R_D$ = 7500 Ω, $R_S$ = 3×10$^6$ Ω, $C_S$ = 2.3×10$^{-19}$ F, $C_D$ = 4.6×10$^{-19}$ F.

The most likely explanation for the gap observed in our experiment is that it arises from effect of exchange-type zero-bias anomaly (ZBA) on single electron tunnelling of Au NPs. Very recently, Bitton, *et al.* reported that in the region, $k_BT << e^2/C$ and $R_t \sim R_q$, the effects of ZBA and CB would coexist in an isolated Au NP [24]. Here we point out that the effect of exchange-type ZBA will result in irregular CB behavior, beyond the description of the orthodox CB theory, of metallic NPs, as shown in Fig. 2. Now we give an overview of the theoretical framework that supports our interpretation of the observed gap. Because of the coexistence of ZBA and CB in the Au NPs, the exchange-type ZBA shifts the first CB peak from $+e/2C$ to 0 (or from $-e/2C$ to 0) and suppresses the single electron tunnelling peak near zero-bias, which results in the gap $\Delta V \sim 2E_C$ [Eq. (2), and final plot in Fig. 2 along with the experimental data].

For the case that $R_t$ and $R_q$ are comparable, CB and ZBA would coexist [24] and the tunnelling current from the source through the Au NP to the drain can be expressed as [9,24]

$$I(V) = G_{as}V - I_0(T,V) - gG_{as}Ve^{-F(T,V)}\cos\left(\frac{2\pi Q_{av}(V)}{e}\right). \quad (1)$$

Here, $G_{as}=1/(R_S+R_D)$, $g = \frac{4.88e^2 R_S R_D}{h(R_S+R_D)} + 11.29$,

$Q_{av}(V) = \frac{C_S R_S - C_D R_D}{R_S + R_D}V + C_g V_g$. In Eq. (1), the first term describes an Ohmic current, the second term reflects a conductance dip at $V \sim 0$ which arises from exchange-type ZBA. The last term represents to CB effect. Differentiating Eq. (1) with the expression of $I_0(T,V)$ and $F(T,V)$, we obtain the differential conductance through the dot as:

$$\frac{G(V)}{G_{as}} = \begin{cases} 1 - \frac{1}{g_D}\left[\ln\left(1+\frac{\hbar^2}{\varepsilon^2 t_C^2}\right) - \frac{2}{1+(\varepsilon t_C/\hbar)^2}\right] - g\exp\left[-\frac{g_D}{2} - \frac{2\pi^2}{e^2}C\left(k_BT + \frac{R_D}{R_S}eV\right)\right] \times \left[\cos\left(2\pi\frac{C_S}{e}V+\phi\right) - \frac{2\pi C_S V}{e}\sin\left(2\pi\frac{C_S}{e}V+\phi\right)\right], & G(V)>0 \\ 0 & G(V)<0 \end{cases} \quad (2)$$

Here, $g_D = \frac{h}{e^2 R_D}$, $t_C = R_D C$, $\phi = 2\pi\frac{C_g}{e}V_g$, and $\varepsilon = \left[(eV)^2 + (2\pi k_B T)^2\right]^{\frac{1}{2}}$. In Eq. (2), we let $G(V)/G_{as} = 0$ once $G(V) < 0$, which corresponds to negative differential conductance. Below we will give a simple justification of this assumption. Theoretically, the negative differential conductance in Eq. (2) arises from the phenomenological theory which describes the CB behavior by the cosine and sine terms [9]. This phenomenological theory is a good approximation if ZBA is dominant and there is small residual CB effect in the asymmetric DBTJ [24]. For the case that the contribution from CB is dominant, the oscillations arising from the cosine and sine terms could be pronounced and result in periodic negative differential conductance along with periodic single electron tunnelling peaks with voltage widths of $e/C$. Experimentally, the negative differential conductance was usually observed in a small bias region, for example, in a STM tip-vertical coupling of two particles-substrate junction [25]. There is no physical reason for the appearance of periodic negative differential conductance. Therefore, it is reasonable to let $G(V)/G_{as} = 0$ once $G(V) < 0$.

In Eq. (2) the second term represents the suppression of the tunnelling conductivity at the Fermi level due to the exchange-type ZBA, and the third term describes the periodic CB oscillations, which show a single electron tunnelling peak at $V = 0$. The competition between effects of ZBA and CB suppresses the tunnelling conductance peak at $V = 0$ and results in a gap $\Delta V \sim 2E_C$ around zero bias. In Fig. 2 the model differential conductance is plotted as a function of voltage bias (red dashed curves), with a best fit of the values of $R_D$, $R_S$, $C_S$, and $C_D$ to the experimental data (black solid curves). In the calculation, we fix $C_D = 2C_S$ with taking into account the parameters of the asymmetric double-barrier tunnel junction in our experiment and obtain $R_D = 7.5$ kΩ $< R_q = 25.8$ kΩ. Obviously, Eq. (2) gives a good description of our experimental results. The obtained result that $R_D < R_q$ further demonstrates that one can observe single electron tunnelling even with the resistance of the junction smaller than $R_q$ [24], which is contrary to the orthodox CB theory [21,22]. For the case $R_D$ (and $R_S$) >> $R_q$, we could observe equaldistant tunnelling peaks as predicted by the orthodox CB theory and the first peaks locates at $+e/2C$ and $-e/2C$ for positive and negative bias, as demonstrated sufficiently in literature [1,4,5]. For the case $R_D$ (or $R_S$) $\sim R_q$, the effects of exchange-type ZBA become important. According to our experiment and analysis, the ZBA shifts the first CB peak from $+e/2C$ to 0 (or from $-e/2C$ to 0) and suppresses the tunnelling conductance at zero bias, which results in a gap $\Delta V \sim 2E_C$ around zero bias, as shown in Fig. 2.

According to Eq. (2), the ratio of $R_D/R_S$ dominates the bias dependent magnitude of the conductance peaks. Two typical differential conductance spectrums of our Au NPs are dividing by a critical ratio $R_D/R_S \sim 0.003$ (if we fix $R_D \sim 7.5$ kΩ, the critical resistance of $R_S$ is about $2.3 \times 10^6$ Ω). For the case $R_D/R_S > 0.003$, the magnitude of the conductance peaks decreases quickly with decreasing the bias from -2 V to 0 V and then almost keeps a constant with increasing the bias from 0 V to 2 V, as shown in Fig. 2(a). For the opposite case that $R_D/R_S < 0.003$, the magnitude of the conductance peaks is almost symmetrical about zero bias, as shown in fig. 2(b). The variation of $R_S$ may arise from the varying of both the contacted resistance and the distance between the Au NP and the substrate.

In conclusion, we studied the single electron tunnelling of isolated ligand-stabilized Au NPs with 1.4 nm in radius by STM and STS at 77K. A novel CB behavior, distinct from that predicted by the orthodox CB theory, with a gap $\Delta V \sim 2E_C$ around zero bias is observed in several different Au NPs. This unexpected result is well explained by taking into account the suppression of the single electron tunnelling conductance at zero bias by the exchange-type ZBA.

This work was supported by the National Natural Science Foundation of China (Grant Nos. 10974019 and 11004010) and the Fundamental Research Funds for the Central Universities.

Email: a: helin@bnu.edu.cn,
       b: jcnie@bnu.edu.cn,
       c: ydli@tsinghua.edu.cn
*The corresponding author.

555